\documentclass[10pt, final, journal, letterpaper, twocolumn]{IEEEtran}
\makeatletter
\def\ps@headings{%
\def\@oddhead{\mbox{}\scriptsize\rightmark \hfil \thepage}%
\def\@evenhead{\scriptsize\thepage \hfil \leftmark\mbox{}}%
\def\@oddfoot{}%
\def\@evenfoot{}}
\makeatother 
\IEEEoverridecommandlockouts

\usepackage{graphicx}
\usepackage{amsmath,amssymb}

\usepackage{algorithm}
\usepackage{algorithmic}
\usepackage{amsmath}
\usepackage{graphics}
\usepackage{epsfig}

\usepackage[numbers,sort&compress]{natbib}
\usepackage{flafter}

\begin{document}
\title{Secure Transmission in Linear Multihop Relaying Networks}
\author{\IEEEauthorblockN{Jianping~Yao,
Xiangyun~Zhou,~\IEEEmembership{Senior Member,~IEEE},
Yuan~Liu,~\IEEEmembership{Member,~IEEE},
and
Suili~Feng,~\IEEEmembership{Member,~IEEE}
}
\thanks
{
J. Yao, Y. Liu, and S. Feng are with School of Electronic and Information Engineering, South China University of Technology, Guangzhou 510641, China (e-mails: yaojp\_scut@qq.com, eeyliu@scut.edu.cn, fengsl@scut.edu.cn).

X. Zhou is with the Research School of Engineering, The Australian National University, Canberra, ACT 0200, Australia (e-mail: xiangyun.zhou@anu.edu.au).
}
}
\maketitle

\begin{abstract}

This paper studies the design and secrecy performance of linear multihop networks, in the presence of randomly distributed eavesdroppers in a large-scale two-dimensional space. Depending on whether there is feedback from the receiver to the transmitter, we study two transmission schemes: on-off transmission (OFT) and non-on-off transmission (NOFT). In the OFT scheme, transmission is suspended if the instantaneous received signal-to-noise ratio (SNR) falls below a given threshold, whereas there is no suspension of transmission in the NOFT scheme. We investigate the optimal design of the linear multiple network in terms of the optimal rate parameters of the wiretap code as well as the optimal number of hops. These design parameters are highly interrelated since more hops reduces the distance of per-hop communication which completely changes the optimal design of the wiretap coding rates. Despite the analytical difficulty, we are able to characterize the optimal designs and the resulting secure transmission throughput in mathematically tractable forms in the high SNR regime. Our numerical results demonstrate that our analytical results obtained in the high SNR regime are accurate at practical SNR values. Hence, these results provide useful guidelines for designing linear multihop networks with targeted physical layer security performance.

\end{abstract}


\IEEEpeerreviewmaketitle

\section{Introduction}

\subsection{Background and Motivation}

With the rise of the Internet of Things (IoT), the need for wireless networks that offer a wide variety of quality-of-service (QoS) features is fast growing. It becomes clear that guaranteeing reliable and secure transmission are two major issues. Due to wireless signal attenuation with transmission distance, cooperative relaying has been considered as an effective method to increase the range and reliability of wireless networks. Several relaying strategies have been adopted in major wireless standards. At the same time, due to the broadcast nature of wireless channels, wireless communication is subject to a wide range of security threats. Traditionally, security risks have been addressed at the upper layers of the wireless network protocol stack. More recently, physical layer security is emerged as a new and complementary security solution that exploits the physical characteristics of the wireless channel from an information-theoretic point of view.  %%%\cite{Bloch2011,Zhou2013}.

For two-hop wireless networks, there exists abundant research publications considering the physical layer security for wireless networks, e.g., \cite{Dong2010,Zheng2015,Cai2014,Mo2012}. Early studies, such as \cite{Dong2010},
investigated how to achieve physical layer security with the conventional relaying schemes like decode and forward (DF) and amplify and forward (AF). New cooperation strategies like cooperative jamming (CJ) were also introduced as secrecy enhancements. When designing the two-hop network, the power allocation and rate adaptation were important parameters to optimize. Relay selection was also an effective way of improving the secrecy performance when multiple potential relays are available. The modeling of the node locations, e.g., for a network having multiple potential relays and multiple eavesdroppers, were done either deterministically or statistically using stochastic geometry tools.

While the aforementioned works focused on two-hop relaying systems, it is worth investigating the secure communication in more elaborate networks, which take more than two hops. However, extending the analysis from two-hop networks to multihop networks is non-trivial, because more hops means that more nodes are involved in the transmission as well as more chances for eavesdropping. In addition, the number of hops becomes a design parameter affecting the end-to-end delay and hence throughput. In this work, we study the interesting but challenging scenario of multihop relaying networks.

\subsection{Related Work and Novelty of Our Study}

Only a few studies have addressed the physical layer security in multihop relaying systems \cite{Saad2012,sheikholeslami2016energy,ghaderi2015minimum,he2013end,lee2015full,yjpjrnl2015,tomasin2014routing,moosavi2016delay}. Among these works, the majority of them addresses the secure routing design with various considerations. For example, the authors in \cite{Saad2012} proposed a tree-formation game to choose secure paths in uplink multihop cellular networks. The authors in \cite{sheikholeslami2016energy,ghaderi2015minimum} considered minimum energy routing in the presence of either multiple malicious jammers or eavesdroppers, to guarantee certain end-to-end performance. The authors in \cite{he2013end,lee2015full} considered the problem of how to communicate securely with the help of untrusted relays and full-duplex jamming relays, respectively. The authors in \cite{yjpjrnl2015} addressed the secure routing problem in multihop wireless networks with half-duplex DF relaying, where the locations of the eavesdroppers were modeled as a homogeneous Poisson point process (PPP).

By reviewing the existing studies on multihop wireless networks, we see that there is some knowledge gap at a fundamental level. Questions like what is the optimal number of hops for a given pair of source and destination is largely an open problem. Although the prior studies on secure routing have somewhat addressed this problem for given network configurations, the focus there was to find the best route for given locations of intermediate nodes, instead of directly analyzing the optimal number of hops when the network allows one to deploy the relays or to select relays from a large pool of available nodes. In addition, it is unclear how the optimal design of secure transmission at each hop is affected by the number of hops.

To the best of our knowledge, only one recently published work in \cite{Nardelli2016Throughput} directly addressed the knowledge gap identified above. Specifically, this work considered a linear multihop network in the presence of randomly distributed eavesdroppers and addressed the question of the optimal number of hops. Deviating from the most common physical-layer-security approach of using wiretap code, the work in \cite{Nardelli2016Throughput} adopted ordinary code that does not provide any level of information-theoretic secrecy. Without wiretap code, the legitimate users have a significantly reduced level of control over the secrecy performance when designing their transmission strategy. In contrast, we consider the use of wiretap code and define secrecy performance from an information-theoretic viewpoint as commonly done in the literature of physical layer security. In this way, the design of the wiretap coding rates has direct impact on both the throughput performance and the secrecy performance. Therefore, the novel contribution of our work is the obtained design guideline on the transmission strategy and number of hops in securing a multihop relaying network with wiretap code protection. As will be discussed, the design guidelines obtained with wiretap code (i.e., this work) and without wiretap code (i.e., \cite{Nardelli2016Throughput}) are very different.

\subsection{Our Approach and Contribution}

In this paper, we study the problem of secure transmission design in a linear multihop wireless network in the presence of randomly distributed eavesdroppers whose locations are modeled using a homogeneous PPP. The relays adopt the randomize-and-forward (RaF) relaying protocol where each relay generates the transmitted codeword independently so that secrecy of individual hops guarantees the secrecy of the entire path \cite{Koyluoglu2012}. The celebrated wiretap code is used to provide the desired physical layer security. We attempt to answer a fundamental question: If the network allows one to deploy equally-spaced relays or to select equally-spaced relays from a large pool of available nodes in a dense network, what is the optimal number of hops and what is the corresponding optimal design of the coding rates for achieving the best physical layer security performance? To answer this question, we formulate a throughput maximization problem with an end-to-end secrecy outage probability constraint. Solving the problem is a non-trivial task because the design parameters are highly interrelated. Having more hops reduces the per-hop communication distance, meaning that a higher rate can be used. On the down side, more hops not only increases the total time for communication but also gives the eavesdroppers more chance to intercept the message. Clearly the encoding rates of the wiretap code need to be carefully designed to achieve the best tradeoff between throughput performance and secrecy performance.

The main contributions of this paper are summarized as follows:
\begin{itemize}
	\item Using a stochastic geometry model for the eavesdroppers' locations, we derive an analytical expression for the end-to-end secrecy outage probability, which is used to measure the secrecy performance of the multihop wireless network.
    \item Depending on whether there is feedback from the receiver to the transmitter, we consider two transmission schemes: on-off transmission (OFT) and non-on-off transmission (NOFT). For both schemes, we solve the throughput maximization problem under a given secrecy outage probability constraint. In particular, the optimal rate parameters of the wiretap code are obtained in mathematically tractable forms.
        \item We obtain further analytical insights on the optimal design parameters and the achievable throughput in the asymptotic high signal-to-noise ratio (SNR) regime. These high SNR results are actually found to be accurate at practical SNR values as verified by numerical results. Regarding the optimal number of hops, our results show that the optimal value is insensitive to the change in operating SNR. On the other hand, the optimal number of hops increases as the density of eavesdroppers increases.
\end{itemize}

It is necessary to compare our results with the ones in \cite{Nardelli2016Throughput} which considered the same network scenario (but with the addition of randomly distributed interferers) and addressed the same question of the optimal number of hops. As described before, the work in \cite{Nardelli2016Throughput} did not consider the use of wiretap code which is a key technique in physical layer security. Under such a framework, the main conclusions in \cite{Nardelli2016Throughput} were (i) a greater number of hops are preferable to a smaller number of hops in any situation; and (ii) imposing a (more stringent) secrecy constraint does not change the maximum achievable throughput. In contrast to \cite{Nardelli2016Throughput}, our framework adopts wiretap code. Consequently, the achievable throughput is clearly a function of the required secrecy constraint. Regarding the optimal number of hops, our finding is certainly not ``the more the merrier''. For example, the optimal number of hops reduces as the eavesdropper's density reduces. To sum up, our definition of secrecy and the considered secure transmission schemes are fundamentally different from \cite{Nardelli2016Throughput}, which leads to very different conclusions on the optimal system design and the resulting performance. The work in \cite{Nardelli2016Throughput} and our work complement each other and give different design guidelines depending on whether wiretap code is used or not.

\begin{table}[t]
\centering
\caption{List of Notation}
\label{tab:List}
\begin{tabular}{ll}
$\alpha$                         &Path loss exponent $\left(2 \leq\alpha\leq 6\right)$\\
${\Phi _{ne}}$                   &Poisson point process of eavesdroppers' location\\
$\lambda_e$                      &Density of ${\Phi _{ne}}$\\
$p$                              &Transmit power\\
$N$                              &Number of hops\\
$L$                              &S-D distance\\
$D_n$                            &Transmission distance of $n$th hop\\
$H_n$                            &Channel fading gain of $n$th hop\\
$S_{ne}$                         &Distance of the eavesdropping channel of $n$th hop\\
$X_{ne}$                         &Channel fading gain of eavesdropping channel of $n$th hop\\
$R_t$                            &Rate of the transmitted codewords\\
$R_s$                            &Rate of the confidential information\\
$R_e$                            &Rate loss for securing the messages against eavesdropping\\
${\mathcal{P}_{t}^{'}}$         &Transmission probability over a single hop\\
${\mathcal{P}_{\textrm{c}}^{'}}$      &Connection probability over a single hop\\
${\mathcal{P}_{\textrm{c}}}$ &End-to-end connection probability over the path\\
${\mathcal{P}_{\textrm{so}}^{'}}$        &Secrecy outage probability over a single hop\\
${\mathcal{P}_{\textrm{so}}}$      &End-to-end secrecy outage probability over the path\\
$\epsilon$                       &Constraint on ${\mathcal{P}_{\textrm{so}}}$\\
$\beta_t$                        &SNR threshold for decoding the message correctly\\
$\beta_e$                        &SNR threshold for secrecy outage\\
$C_{{ne}}^\textrm{max}$    &Maximum capacity of eavesdropping channels of $n$th hop\\
${{\tt SNR}_{{ne}}^\textrm{max}}$&Maximum received SNR at eavesdroppers of $n$th hop\\
$\mathbb{U}$                     &Secure transmission throughput\\
$\mathcal{P}\left(\cdot\right)$ &Probability operator\\
$\mathbb{E}\left(\cdot\right)$  &Expectation operator\\
$\Gamma\left(\cdot\right)$      &Gamma function\\
$\mathbb{W}_0\left(\cdot\right)$&Principal branch of Lambert W function
\end{tabular}
\end{table}

The remainder of this paper is organized as follows. In Section \uppercase\expandafter{\romannumeral2}, the system model and performance metric are described. In Section \uppercase\expandafter{\romannumeral3}, the secrecy performance and secure transmission design for a multihop path are investigated. In Section \uppercase\expandafter{\romannumeral4}, the numerical results are presented. Finally, the conclusion is drawn in Section \uppercase\expandafter{\romannumeral5}. Table I summarizes the list of notation used in this paper.

\begin{figure}[t]
\centering
  \includegraphics[width=8.0cm]{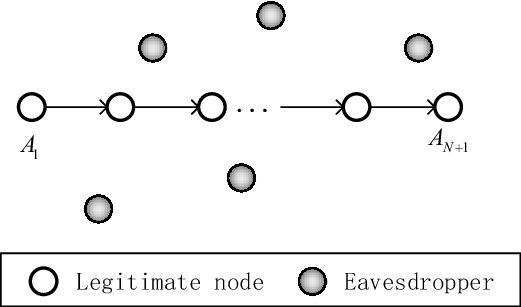}\\
  \caption{An illustration of a linear multihop relaying network surrounded by several eavesdroppers. The confidential message is transmitted from the source node $A_1$ to the destination $A_{N+1}$ with the help of relays $\left(\{{A_i}\},i = 2,\ldots,N\right)$, and at the same time the eavesdroppers attempt to intercept the transmitted message.}\label{fig:system_model}
\end{figure}

\section{System Model and Performance Metric}
\subsection{System Model}
We consider a linear $N$-hop wireless relay network of length $L$ as shown in Fig. \ref{fig:system_model}, consisting of a source node $A_1$, a destination node $A_{N+1}$ and $N-1$ relay nodes $\left(\{{A_i}\},i = 2,\ldots,N\right)$, which is exposed to a set of randomly deployed eavesdroppers over a large two-dimensional area. We model the locations of the eavesdroppers in each hop as an independent homogeneous PPP with intensity ${\lambda_e}$ denoted by ${\Phi _{ne}} \left(n = 1,\ldots,N\right)$. The eavesdroppers are non-cooperative, so they must decode the messages individually. All nodes are equipped with one omni-directional antenna and, hence, cannot transmit and receive signals simultaneously which is referred to as ``half duplex''. Furthermore, the transmission is performed via time-division which guarantees that there is no interference between the different hops. Node $A_{n+1}$ only receives the signal transmitted by its adjacent node $A_{n}$.

In this paper, we assumed that the relays are equidistant with each other as in \cite{waqar2014energy,stamatiou2014delay,sikora2006bandwidth}. Such a model is mathematically tractable to investigate the impact of the number of hops and hence, is well-adopted in the literature in studying multihop network. It can be used to approximate the performance of a dense network, where we can pick approximately equidistant nodes as relays. We can also find it in practical, e.g., the communication between the electrical equipments in smart grid, or the communication between road-side units placed along the road and railway.

All the channels are modeled by large-scale attenuation with path loss exponent $\alpha$ along with small-scale Rayleigh fading. We consider the non-singular (bounded) path loss model $\frac{1}{1+r^\alpha}$, where $r$ denotes the communication distance \cite{guan2016transmit,guo2015asymptotic,giacomelli2011outage,ganti2012spatial,zaidi2015correction}. The corresponding channel power gains are independent and exponentially distributed with unit mean. The noise at each node is assumed to be complex additive white Gaussian with zero mean and variance one. The instantaneous received SNR at the legitimate node $A_{n+1}$ and eavesdropper (at position) $e$ in ${\Phi _{ne}}$ can be respectively given as
\begin{align}
{{\tt SNR}_{n}} \triangleq \frac{{p}{H_{n}}}{{|{D_{n}}|}^{\alpha}+1},
\label{snr_l}
\end{align}
\begin{align}
{{\tt SNR}_{{{ne}}}} \triangleq \frac{{p}{S_{ne}}}{{|{X_{ne}}|}^{\alpha}+1},
\label{snr_e}
\end{align}
where ${p}$ denotes the transmit power at the legitimate node ${A_n}$ assumed the same in each hop; ${H_{{n}}}$ and $|{D_{{n}}}|$ are the channel power fading gain and the distance between the $n$th link ${{A_n}{A_{n + 1}}}$, respectively; ${S_{{n}{e}}}$ and $|{X_{{n}{e}}}|$ are the channel power fading gain and the distance between the legitimate node $A_n$ and eavesdropper $e$, respectively. Since the distance for each hop is the same between the legitimate nodes, the statistics of ${{\tt SNR}_{n}}$ is the same for all $n$. Note that we have normalized the receiver noise power to be one, which means that the parameter $p$ in fact represents the transmitter-side SNR instead of the actual transmit power.

We assume non-cooperative eavesdroppers.  Then the maximum received SNR, ${{\tt SNR}_{{ne}}^\textrm{max}}$, at eavesdroppers  from the legitimate node ${A_n}$ is equivalent to ${\mathop {\max }\limits_{{e} \in {\Phi _{ne}}} \left\{ {{\tt SNR}_{{n}{e}}} \right\}}$, where the maximization operation means the selection of the eavesdropper which has the strongest received signal.

\subsection{Transmission Schemes}
We consider the well-known Wyner's encoding scheme \cite{Wyner1975}. The transmitters encode the message using two rate parameters, namely, the rate of the transmitted codewords $R_t$ and the rate of the confidential information $R_s$. The rate difference $R_e \triangleq R_t - R_s$ represents the rate loss for transmitting the message securely against eavesdropping which reflects the ability of securing the message transmission against eavesdropping \cite{zhou2013rethinking}. We assume that the intermediate relay nodes use the RaF relaying protocol which is specifically designed from the viewpoint of physical layer security. RaF relaying deviates from the widely-used DF relaying in the way that the relays use independent codewords to add independent randomization in each hop when re-encoding the received signal \cite{Koyluoglu2012}. We consider fixed-rate transmission, hence $R_t$ and $R_s$ are designed offline. This is a commonly-adopted practical assumption as its implementation is based on long-term channel statistics and does not require instantaneous feedback of the full channel state information (CSI) for every hop. For reducing the risk of eavesdropping, we assume that there is no retransmission in each hop. Depending on whether there is feedback from the receiver to the transmitter, we can have the following two transmission schemes: OFT and NOFT, as described below.

For a given transmission rate $R_t$, the receiver is able to decode the transmission if the instantaneous received SNR exceeds a threshold $\beta_t$, where $\beta_t = 2^{R_t} - 1$. Assuming the receiver has perfect channel knowledge (e.g., obtained from pilot transmissions), it is possible for the receiver to feed back to the transmitter some information about its channel status. In the OFT scheme \cite{zhou2013rethinking}, we assume that the receiver uses a one-bit feedback (as opposed to full CSI feedback) to inform the transmitter whether the instantaneous received SNR exceeds $\beta_t$ or not. Transmission is suspended if the received SNR falls below $\beta_t$. Then, the transmission probability of the $n$th hop can be defined as
\begin{align}
{\mathcal{P}_{t}^{{'}}} \triangleq {\mathcal{P}}\left ({\tt SNR}_{{n}}> \beta_t\right ).
\label{P_t_OFT_1}
\end{align}

Since the statistics of ${{\tt SNR}_{n}}$ is the same for all $n$, ${\mathcal{P}_{t}^{{'}}}$ is the same for all hops. Although the transmission suspension causes delay as reflected in the transmission probability, it guarantees successful connection/decoding for each hop whenever transmission happens.

If there is no feedback, i.e., in the NOFT scheme, there is no suspension of transmission, and hence, the transmission probability is ${\mathcal{P}_{t}^{'}} = 1$. However, the receiver may not be able to decode. Specifically, the connection probability (i.e., the probability that the receiver is able to decode the message) of the $n$th hop is defined as
\begin{align}
{\mathcal{P}_{c}^{'}} \triangleq {\mathcal{P}}\left ({\tt SNR}_{{n}} > \beta_t\right).
\label{P_c_NOFT_1}
\end{align}

Hence, the end-to-end connection probability of the path in the NOFT scheme is given by
\begin{align}
{\mathcal{P}_{\textrm{c}}} \triangleq \mathcal{P}\left ( \min\limits_{n = 1,\ldots,N}\left \{{\tt SNR}_{{n}}\right \} > \beta_t\right ).
\label{P_c_path_1}
\end{align}

For any given $R_t$ and $R_s$, the secrecy outage probability of the $n$th hop is defined as \cite{zhou2013rethinking}
\begin{align}
{\mathcal{P}_{\textrm{so}}^{'}} = {\mathcal{P}}\left (C_{\textrm{ne}}^\textrm{max}> R_t-R_s\right)= {\mathcal{P}}\left ({{\tt SNR}_{{ne}}^\textrm{max}} > \beta_e\right),
\label{P_so_1}
\end{align}
where $C_{\textrm{ne}}^\textrm{max}$ is the maximum capacity of the eavesdroppers' channels in the $n$th hop and $\beta_e = 2^{R_e}-1$.

Because the relays apply the RaF protocol, the source and relays use different codewords to transmit the secret message. According to \cite{Koyluoglu2012}, the message is secure when every hop in the path is secure. Hence, the end-to-end secrecy outage probability of the path can be expressed as
\begin{align}
{\mathcal{P}_{\textrm{so}}} \triangleq \mathcal{P}\left ( \max\limits_{n = 1,\ldots,N}\left \{{{\tt SNR}_{{ne}}^\textrm{max}}\right \}> \beta_e\right ).
\label{P_so_path_1}
\end{align}

Furthermore, we assume that the point processes ${\Phi _{ne}} \left(n = 1,\ldots,N\right)$ representing the eavesdroppers' locations in different hops are independent, which is a worse case from the view of security compared to the scenario where the eavesdroppers' locations are fixed during all hops. In the next section, we will prove that the secrecy outage probability under the independent point-processes assumption is strictly higher than the secrecy outage probability under the fixed eavesdropper-locations assumption. Because the legitimate nodes have little knowledge on the eavesdroppers' locations or their mobility, it is best to consider a worse case scenario from the security point of view.

Since the statistics of the legitimate link is the same and the fading gains are independent in each hop, the end-to-end connection probability in the NOFT scheme defined in (\ref{P_c_path_1}) is equivalent to
\begin{align}
{\mathcal{P}_{\textrm{c}}} =\left({\mathcal{P}_{\textrm{c}}^{'}}\right)^N.
\label{P_c_path_2}
\end{align}

Clearly the end-to-end connection probability in the OFT scheme is 1, but this is at the price of the reduced transmission probability as described in (\ref{P_t_OFT_1}).

For both NOFT and OFT schemes, the end-to-end secrecy outage probability defined in (\ref{P_so_path_1}) is equivalent to
\begin{align}
{\mathcal{P}_{\textrm{so}}} =1- \left(1-{\mathcal{P}_{\textrm{so}}^{'}}\right)^N.
\label{P_so_path_2}
\end{align}

\subsection{Secure Transmission Throughput}
Secure transmission throughput characterizes the spectral efficiency of secure communication in a given multihop path for a source-destination pair of nodes which is defined as the average end-to-end rate of the transmission of confidential messages that can be sustained reliably, normalized by the total transmission time:
\begin{align}
\mathbb{U} = \frac{{\mathcal{P}_t^{'}}{\mathcal{P}_c}{R_s}}{N},
\label{throuhput_1}
\end{align}
where $\frac{{N}}{{\mathcal{P}_t^{'}}}$ is the total expected transmission time (in slots) for transmitting a confidential message from source to destination which includes the transmission time and waiting time along the path; ${\mathcal{P}_c}$ represents the probability that a confidential message is transmitted correctly from the source to the destination over the path. Secure transmission throughput shows the dependence of the network spectral efficiency on the key system parameters, e.g., the transmission rate parameters, the number of hops, the transmit power, and the density of eavesdroppers.

As explained before, in the NOFT scheme, the transmission probability of a single hop ${\mathcal{P}_t^{'}}=1$ and the total expected transmission time is $N$. Then, secure transmission throughput can be rewritten as $\mathbb{U} = \frac{{\mathcal{P}_c}{R_s}}{N}$, which is the same as the definition of secure transmission throughput in \cite{Zhou2011,andrews2010random,Chen2012Upper}. On the other hand, in the OFT scheme, the connection probability of the path ${\mathcal{P}_{\textrm{c}}}=1$ and the total expected transmission time is $\frac{{N}}{{\mathcal{P}_t^{'}}}$. Then, secure transmission throughput can be rewritten as $\mathbb{U} = \frac{{\mathcal{P}_t^{'}}{R_s}}{N}$, which is the same as the definition of secure transmission throughput in \cite{zhou2013rethinking}. Hence, our throughput metric is consistent with existing work in this area. We will use the secure transmission throughput as the main performance metric in this paper. Note that the throughput definition in (\ref{throuhput_1}) alone does not directly describe the secrecy performance in terms of the secrecy outage probability. We use the end-to-end secrecy outage probability in (\ref{P_so_path_1}) to directly quantify the level of security for the multihop communication.

\section{Secure Transmission Design in a Multihop Path}
In this section, we analytically study the secrecy outage performance of a given multihop path of a source-destination pair of nodes. Then, for both OFT and NOFT schemes, we consider the secure transmission design to maximize secure transmission throughput under the secrecy outage constraint.

\subsection{Secrecy Performance and Throughput Maximization Problem}
In this subsection, we derive an explicit expression of the end-to-end secrecy outage probability. Then, we formulate the throughput maximization problem by the joint design of the number of hops, the rate of the transmitted codewords and the rate of the confidential information.

\newtheorem{theorem}{Theorem}
\begin{theorem}\label{Theorem_P_so}
The end-to-end secrecy outage probability of the path for both NOFT and OFT schemes is given by
\begin{align}
{\mathcal{P}_{\textrm{so}}} = 1-\exp\left [ -{N}{K_1}\left (\frac{\beta_e}{p}\right)^{-\frac{2}{\alpha}}\exp\left [ -\frac{\beta_e}{p} \right ]\right ],
\label{P_so_path_3}
\end{align}
where $K_1 = \pi\lambda_e \Gamma\left (\frac{2}{\alpha}+1 \right)$ and $\Gamma ( \cdot )$ is the gamma function.

\begin{IEEEproof}
See Appendix \ref{appendices_Theorem_P_so}.
\end{IEEEproof}
\end{theorem}
\vspace{3ex}
%
%\begin{figure}[t]
%\centering
%  \includegraphics[width=8.6cm]{M_RF_N.eps}\\
%  \caption{The end-to-end secrecy outage probability ${\mathcal{P}_{\textrm{so}}}$. We consider a scenario that the source node is located at the origin and the destination node is located at $\left(10\textrm{m},0\right)$. The eavesdroppers are randomly distributed in the entire network of size
%$2000\textrm{m} \times 2000\textrm{m}$. The system parameters are $L = 10\textrm{m}$, $\alpha = 3$, $\beta_e = 14.9 \textrm{dB}$, $p = 30 \textrm{dB}$. The analytical results are based on (\ref{P_so_path_3}). For Monte Carlo simulation, it takes 10000 simulation runs to obtain the results. }\label{fig:M_DF_C}
%\end{figure}
%As demonstrated in Fig. 2, our analytical results match with the Monte Carlo simulation results, which validates our analysis. In the following, we will adopt (\ref{P_so_path_3}) as the metric to character the secrecy performance.

Now we revisit the assumption on the eavesdroppers' location made in Section \uppercase\expandafter{\romannumeral2}-A and provide a justification for it. In particular, we have assumed that the eavesdroppers' locations change independently from hop to hop. Of course, this assumption does not accurately reflect the realistic locations of eavesdroppers, unless they have extremely high mobility. Without any knowledge of the eavesdroppers' locations and mobility, however, we have to make some assumption in order to assess the network performance. In the following corollary, we show that the assumption adopted in this work is more robust than assuming that the eavesdropper locations are fixed over all hops (i.e., stationary eavesdroppers).

\newtheorem{corollary}{Corollary}
\begin{corollary}\label{coro_fix}
The secrecy outage probability under the independent point-processes assumption is strictly higher than the secrecy outage probability under the fixed eavesdropper-locations assumption.
\begin{IEEEproof}
See Appendix \ref{appendices_coro_fix}.
\end{IEEEproof}
\end{corollary}
\vspace{3ex}

The result in Corollary \ref{coro_fix} agrees with intuition because that the eavesdroppers under the independent point-processes assumption have more degrees of freedom than the eavesdroppers under the fixed locations assumption, that is, more variation in the locations gives eavesdroppers more chance to cause secrecy problem to at least one of the hops. Hence, when not knowing the exact locations or mobility of the eavesdroppers, it is more appropriate for the designers of the legitimate network to consider a worse-case scenario, i.e., the independent point-processes assumption adopted in this work.

With the end-to-end secrecy outage probability derived in Theorem 1 to quantify the secrecy performance, we now formulate a network design problem of maximizing the secure transmission throughput subject to a given secrecy requirement:
\begin{align*}
\underset{R_t,R_s,N}{\max}\;\; \mathbb{U} = \frac{{\mathcal{P}_t^{'}}{\mathcal{P}_c}{R_s}}{N}, \quad s.t. \;\; {\mathcal{P}_{\textrm{so}}}\leq \epsilon,
\end{align*}
where $\epsilon\in \left[0,1\right]$ represents the minimum security requirement. The controllable design parameters are the rate of the transmitted codewords $R_t$, the rate of the confidential information $R_s$, and the number of hops $N$ of the path.

\subsection{On-Off Transmission Scheme}
We first consider the OFT scheme, i.e., transmission occurs whenever the received SNR at the legitimate node exceeds the SNR threshold $\beta_t$. Hence, for any transmitted message, the legitimate receiver is able to decode correctly, i.e., the end-to-end connection probability of the path ${\mathcal{P}_{\textrm{c}}} = 1$.

According to (\ref{snr_l}) and (\ref{P_t_OFT_1}), the transmission probability can be computed as
\begin{align}
{\mathcal{P}_{t}^{'}} = {\mathcal{P}}\left (\frac{{p}{H_{{n}}}}{{\left({ \frac{{L}}{N}}\right)}^{\alpha}+1}> \beta_t\right ) = \exp\left [ \frac{-\beta_t\left [ \left ( \frac{L}{N} \right )^\alpha +1 \right ]}{p}\right].
\label{P_t_onoff_2}
\end{align}

Now, we consider the design problem of maximizing secure transmission throughput by the joint design of $R_t$, $R_s$ and $N$, expressed as
\begin{subequations}
\begin{align}
\label{P1_a}\textbf{P1}:~~ \underset{R_t,R_s,N}{\max}\quad &\mathbb{U} = \frac{{\mathcal{P}_t^{'}}\left(R_t,N\right){R_s}}{N}, \\
\label{P1_b}s.t.\quad &{\mathcal{P}_{\textrm{so}}}\left(R_t,R_s,N\right)\leq \epsilon,\\
\label{P1_c}&R_t\geq R_s >0,\\
\label{P1_d}&N\geq 1,
\end{align}
\end{subequations}
where we have explicitly shown the dependence of ${\mathcal{P}_{\textrm{so}}}$ and ${\mathcal{P}_t^{'}}$ on $R_t$, $R_s$ and $N$.

\subsubsection{The Optimal Rate Parameters $R_t$ and $R_s$ for Fixed Hop-Count $N$}
To solve the above optimization problem $\textbf{P1}$, we first consider the sub-problem that we get the optimal rate parameters $R_t$ and $R_s$ for fixed hop-count $N$. This sub-problem has its important physical meaning: how to optimally design the encoding rates for a given network setting.

Since the constraint (\ref{P1_b}) is satisfied only when the constraint (\ref{P1_c}) is satisfied, the constraint (\ref{P1_c}) can be simplified as $R_s >0$. Since ${\mathcal{P}_t^{'}}$ is independent of $R_s$ and ${\mathcal{P}_{\textrm{so}}}$ is an increasing function of $R_s$, it is optimal to maximize ${\mathcal{P}_{\textrm{so}}}$ in order to maximize $R_s$ in $\mathbb{U}$. Hence, we can obtain the optimal ${\mathcal{P}_{\textrm{so}}}$ as
\begin{align}
{\mathcal{P}_{\textrm{so}}}\left(R_t,R_s,N\right) = \epsilon.
\label{P_so_path_4}
\end{align}
The value of $\beta_e$ that satisfies the above equality is given as
\begin{align}
{\beta_e} = \frac{2{p}}{\alpha}\mathbb{W}_0\left (\frac{\alpha}{2} \left [\frac{\ln{\frac{1}{1-\epsilon}}} {{N}{K_1}} \right ]^{-\frac{\alpha}{2}} \right ),
\label{Beta_e}
\end{align}
where $\mathbb{W}_0\left (\cdot\right )$ is the principal branch of Lambert W function. Then, from (\ref{Beta_e}), we can derive
\begin{align}
R_e = {R_t - R_s}
= \log_2\left [\frac{2{p}}{\alpha}\mathbb{W}_0\left (\frac{\alpha}{2} \left [\frac{\ln{\frac{1}{1-\epsilon}}} {{N}{K_1}} \right ]^{-\frac{\alpha}{2}} \right ) +1 \right ].
\label{R_e}
\end{align}

Under this condition, we can reformulate the optimization problem as
\begin{subequations}
\begin{align}
\label{P2_a}\textbf{P1}':~ \underset{R_t}{\max}\quad &\mathbb{U} = \frac{{\left ( {R_t - R_e} \right )}{\exp\left [ -{K_2}{\left ( 2^{R_t}-1 \right )}\right]}}{N}, \\
\label{P2_b}s.t.\quad &R_t > R_e,
\end{align}
\end{subequations}
where $K_2 =  \frac{\left [ \left ( \frac{L}{N} \right )^\alpha +1 \right ]}{p}$; $R_e$ is a function of $N$, whose explicit expression can be found in (\ref{R_e}).

\begin{theorem}\label{Theorem_R_t_R_s}
Secure transmission throughput $\mathbb{U}$ is a quasi-concave function of the rate of the transmitted codewords $R_t$. Then, the optimal value of $R_t$ and $R_s$ to maximize $\mathbb{U}$ are given as
\begin{align}
R_t^* = R_e + \frac{1}{\ln2}\mathbb{W}_0\left (\frac{2^{-R_e}}{K_2} \right ),
\label{R_t}
\end{align}
and
\begin{align}
R_s^* = \frac{1}{\ln2}\mathbb{W}_0\left (\frac{2^{-R_e}}{K_2} \right ).
\label{R_s}
\end{align}

\begin{IEEEproof}
See Appendix \ref{appendices_Theorem_R_t_R_s}.
\end{IEEEproof}
\end{theorem}
\vspace{3ex}

\begin{corollary}\label{coro_N}
The optimal rate of the transmitted codewords $R_t^*$ is an increasing function of the number of hop $N$. And when $N$ goes to infinity, the optimal rate of the confidential information $R_s^*$ approaches $0$.

\begin{IEEEproof}
See Appendix \ref{appendices_coro_N}.
\end{IEEEproof}
\end{corollary}
\vspace{1ex}

The result in Corollary \ref{coro_N} clearly says that a network with too many hops will lead to no secure throughput, since the drawback of causing more chances for eavesdropping outweighs the benefit of allowing more randomness in the code, i.e. $R_s^*$ goes to 0. Hence, we definitely expect that the optimal number of hops to be finite.

\begin{corollary}\label{coro_p}
As ${p}$ grows to infinity, the optimal value of $R_t$ goes to infinity and the optimal values of $R_s$ and $\mathbb{U}$ converge to constants, given as
\begin{align}
R_s^* = \frac{1}{\ln2}\mathbb{W}_0\left (\frac{1}{K_3} \right ),
\label{R_s2}
\end{align}
and
\begin{align}
\mathbb{U}^* =&\frac{{1}}{{N}\ln2}\mathbb{W}_0\left (\frac{1}{{K_3}} \right )\exp\left [ -{K_3}\exp\left [\mathbb{W}_0\left (\frac{1}{K_3} \right ) \right] \right],
\label{throuhput_2}
\end{align}
where $K_3 = \frac{2}{\alpha}{\left [ \left ( \frac{L}{N} \right )^\alpha +1 \right ]}\mathbb{W}_0\left (\frac{\alpha}{2} \left [\frac{\ln{\frac{1}{1-\epsilon}}} {{N}{K_1}} \right ]^{-\frac{\alpha}{2}} \right )$.
\begin{IEEEproof}
See Appendix \ref{appendices_coro_p}.
\end{IEEEproof}
\end{corollary}
\vspace{1ex}

If there is no secrecy requirement for the system, increasing the transmit power can always increase the throughput. However, for the system with secrecy requirement, the improvement of the throughput tends to zero as the transmit power grows to infinity since increasing power benefits both the legitimate and eavesdropping channels.

\subsubsection{The Optimal Hop-Count $N$}
With the obtained explicit expressions of $R_t^*$ and $R_s^*$, in the following, we study how to design the number of hops $N$ of the path. Replacing $R_t$ with (\ref{R_t}), then the optimization problem $\textbf{P1}'$ can be rewritten as:
\begin{subequations}
\begin{align}
\nonumber\textbf{P1}'':~~ \label{P3_a} \underset{N}{\max}&~~ \mathbb{U} =\frac{{ {\mathbb{W}_0\left (\frac{2^{-R_e}}{K_2} \right )}}}{{N}{\ln2}}\\
&~~~~\times{\exp\left [ -{K_2}{\left ( 2^{R_e + \frac{1}{\ln2}\mathbb{W}_0\left (\frac{2^{-R_e}}{K_2} \right )}-1 \right )}\right]},\\
\label{P3_b}s.t.&\quad N\geq 1,
\end{align}
\end{subequations}
where $R_e$ and $K_2$ are functions of $N$, whose explicit expressions can be found earlier. We can see that the objective function is a complicated function of the argument $N$, which makes the optimization problem $\textbf{P1}''$ difficult to be solved. We cannot obtain an explicit expression of the optimal value of $N$, however, this problem can be solved numerically. Note that $N$ is an integer and the feasible range of $N$ is typically small in practical networks. Therefore, it is of minimal complexity to numerically optimize $N$. We will present numerical results on the optimal $N$ in Section \uppercase\expandafter{\romannumeral4}.

\subsection{Non-On-Off Transmission Scheme}
We now consider the NOFT scheme. Since there is no feedback of the instantaneous SNR from the receiver to the transmitter, there is no suspension of transmission. Hence, each hop transmits the message instantly without waiting, i.e., the transmission probability of the $n$th hop ${\mathcal{P}_{t}^{'}} = 1$.

According to (\ref{snr_l}) and (\ref{P_c_NOFT_1}), the connection probability can be computed as
\begin{align}
{\mathcal{P}_{c}^{'}} = {\mathcal{P}}\left (\frac{{p}{H_{{n}}}}{{\left({ \frac{{L}}{N}}\right)}^{\alpha}+1}> \beta_t\right ) = \exp\left [ \frac{-\beta_t\left [ \left ( \frac{L}{N} \right )^\alpha +1 \right ]}{p}\right].
\label{P_c_fi_final}
\end{align}

Replacing ${\mathcal{P}_{\textrm{c}}^{'}}$ with (\ref{P_c_fi_final}) into (\ref{P_c_path_2}), the end-to-end connection probability of the path can be computed as
\begin{align}
{\mathcal{P}_{c}} = \exp\left [ \frac{-N\beta_t\left [ \left ( \frac{L}{N} \right )^\alpha +1 \right ]}{p}\right].
\label{P_c_path_fi_final}
\end{align}

Now, we consider the design problem of maximizing secure transmission throughput by the joint design of $R_t$, $R_s$ and $N$, expressed as
\begin{subequations}
\begin{align}
\label{P6_a}\textbf{P2}:~~ \underset{R_t,R_s,N}{\max}\quad &\mathbb{U} = \frac{{\mathcal{P}_c}\left(R_t,N\right){R_s}}{N}, \\
\label{P6_b}s.t.\quad &{\mathcal{P}_{\textrm{so}}}\left(R_t,R_s,N\right)\leq \epsilon,\\
\label{P6_c}&R_t\geq R_s >0,\\
\label{P6_d}&N\geq 1,
\end{align}
\end{subequations}
where we have explicitly shown the dependence of ${\mathcal{P}_{\textrm{so}}}$ and ${\mathcal{P}_c}$ on $R_t$, $R_s$ and $N$.

\subsubsection{The Optimal Rate Parameters $R_t$ and $R_s$ for Fixed Hop-Count $N$}
To solve the above optimization problem $\textbf{P2}$, we first consider the sub-problem that we get the optimal rate parameters $R_t$ and $R_s$ for fixed hop-count $N$. As explained before, this sub-problem has its important physical meaning: how to optimally design the encoding rates for a given network setting.

Similar to the OFT case, the optimal design needs to satisfy (\ref{P_so_path_4}). Then, the optimal values of $R_t$ and $R_s$  should satisfy (\ref{R_e}).

Under this condition, we can reformulate the optimization problem $\textbf{P2}$ as
\begin{subequations}
\begin{align}
\label{P2_2a}\textbf{P2}':~ \underset{R_t}{\max}\quad &\mathbb{U} = \frac{{\left ( {R_t - R_e} \right )}{\exp\left [ -{K_4}{\left ( 2^{R_t}-1 \right )}\right]}}{N}, \\
\label{P2_2b}s.t.\quad &R_t > R_e,
\end{align}
\end{subequations}
where $K_4 =  \frac{N\left [ \left ( \frac{L}{N} \right )^\alpha +1 \right ]}{p}$; $R_e$ is a function of $N$, whose explicit expression can be found in (\ref{R_e}).

\begin{theorem}\label{Theorem_R_t_R_s_NOFT}
Secure transmission throughput $\mathbb{U}$ is a quasi-concave function of the rate of the transmitted codewords $R_t$. Then, the optimal value of $R_t$ and $R_s$ to maximize $\mathbb{U}$ are given as
\begin{align}
R_t^* = R_e + \frac{1}{\ln2}\mathbb{W}_0\left (\frac{2^{-R_e}}{K_4} \right ),
\label{R_t_OFT_final}
\end{align}
and
\begin{align}
R_s^* = \frac{1}{\ln2}\mathbb{W}_0\left (\frac{2^{-R_e}}{K_4} \right ).
\label{R_s_OFT_final}
\end{align}

\begin{IEEEproof}
See Appendix \ref{appendices_Theorem_R_t_R_s_NOFT}.
\end{IEEEproof}
\end{theorem}
\vspace{3ex}

Comparing with the optimal coding rate parameters for the OFT scheme in Theorem \ref{Theorem_R_t_R_s}, we see that the expressions of $R_t^*$ and $R_s^*$ are very similar between the OFT scheme and the OFT scheme. The only but important difference is between the parameter $K_2$ and $K_4$. Specifically, $K_4 = N K_2$.

\begin{corollary}\label{coro_p_NOFT}
As ${p}$ grows to infinity, the optimal value of $R_t$ goes to infinity and the optimal values of $R_s$ and $\mathbb{U}$ converge to constants, given as
\begin{align}
R_s^* = \frac{1}{\ln2}\mathbb{W}_0\left (\frac{1}{K_5} \right ),
\label{R_s2_NOFT}
\end{align}
and
\begin{align}
\mathbb{U}^* =&\frac{{1}}{{N}\ln2}\mathbb{W}_0\left (\frac{1}{{K_5}} \right )\exp\left [ -{K_5}\exp\left [\mathbb{W}_0\left (\frac{1}{K_5} \right ) \right] \right],
\label{throuhput_2_NOFT}
\end{align}
where $K_5 = \frac{2N}{\alpha}{\left [ \left ( \frac{L}{N} \right )^\alpha +1 \right ]}\mathbb{W}_0\left (\frac{\alpha}{2} \left [\frac{\ln{\frac{1}{1-\epsilon}}} {{N}{K_1}} \right ]^{-\frac{\alpha}{2}} \right )$.
\begin{IEEEproof}
See Appendix \ref{appendices_coro_p_NOFT}.
\end{IEEEproof}
\end{corollary}
\vspace{1ex}

Again, we see that the throughput improvement from increasing the transmit power vanishes as the transmit power goes large.

\subsubsection{The Optimal Hop-Count $N$}
With the obtained explicit expressions of $R_t^*$ and $R_s^*$, in the following, we study how to design the number of hops $N$ of the path. Replacing $R_t$ with (\ref{R_t_OFT_final}), then the optimization problem $\textbf{P2}'$ can be rewritten as:
\begin{subequations}
\begin{align}
\nonumber\textbf{P2}'':~ \label{P7_a} \underset{N}{\max}&~ \mathbb{U} =\frac{{ {\mathbb{W}_0\left (\frac{2^{-R_e}}{K_4} \right )}}}{{N}{\ln2}}\\
&~~~\times{\exp\left [ -{K_4}{\left ( 2^{R_e + \frac{1}{\ln2}\mathbb{W}_0\left (\frac{2^{-R_e}}{K_4} \right )}-1 \right )}\right]},\\
\label{P7_b}s.t.&\quad N\geq 1,
\end{align}
\end{subequations}
where $R_e$ and $K_4$ are functions of $N$, whose explicit expressions can be found earlier. Again, this problem is difficult to be solved analytically. Nevertheless, it is of minimal complexity to numerically find the optimal $N$ as explained before.

\section{Numerical Results}
In this section, we present numerical results on secure transmission throughput and evaluate how different system parameters impact the secure transmission design. We consider a multihop wireless network in which legitimate nodes are placed uniformly on a line in the center of the network. The source node is placed at the origin and the destination is located at $\left(50\textrm{m},0\right)$. The eavesdroppers are randomly distributed according to a uniform distribution in the entire network of size $2000\textrm{m} \times 2000\textrm{m}$. The eavesdroppers' distribution is independent from hop to hop. Unless otherwise stated, we use the following settings to obtain the numerical results: $L=50\textrm{m}$, $\epsilon=0.05$, $\lambda_e = 10^{-5}$, $\alpha=3$, $p = 100$dB (which corresponds to a practical (sensor-like device) scenario with transmit power of 0dBm and a receiver noise power of -100dBm). Note that although $p$ is said to denote the transmit power, it actually presents the transmitter-side SNR due to the normalization in the receiver noise power.

\subsection{Performance of Secure Transmission}
We first show the impact of different system parameters on the design of the encoding rates as well as the secure transmission throughput.

\begin{figure}[!t]
\centering
  \includegraphics[width=8.0cm]{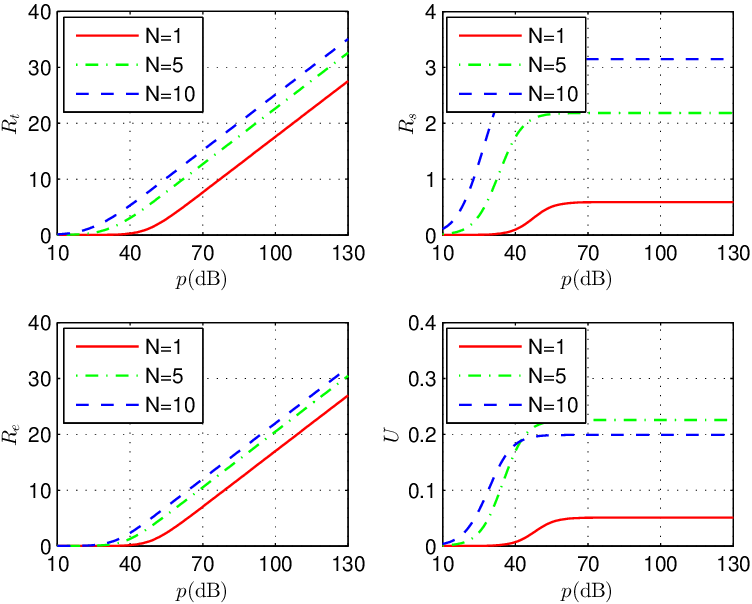}\\
  \caption{Performance of secure transmission under fixed hop-count $N$ versus the transmit power $p$ for the case of OFT scheme. The system parameters are $L = 50\textrm{m}$, $\epsilon = 0.05$, $\alpha= 3$, $\lambda_e = 10^{-5}$.}
\label{fig:OFT_P_0_fixN}
\end{figure}
Fig. \ref{fig:OFT_P_0_fixN} presents the secrecy performance of the system under fixed hop-count $N$ versus the transmit power $p$ for the case of OFT scheme. As $p$ increases, the received power at eavesdroppers increases as well. To maintain the same level of secrecy, the legitimate nodes need to increase the randomness in the wiretap code, i.e., $R_e$, which indirectly requires an increase in $R_t$. This explains the trend seen in the two sub-figures on the left. Also, we see that the slopes of $R_t$ and $R_e$ are the same as $p$ increases above 70dB, which explains why $R_s = R_t - R_e$ is a constant for $p$ above 70dB, hence a constant throughput $U$, which validates our analysis in Proposition \ref{coro_p}. Comparing the throughput with different number of hops, we see that the network with 5 hops performs best among the three different choices of hop-count, indicating that there is an optimal number of hops.

\begin{figure}[!t]
\centering
  \includegraphics[width=8.0cm]{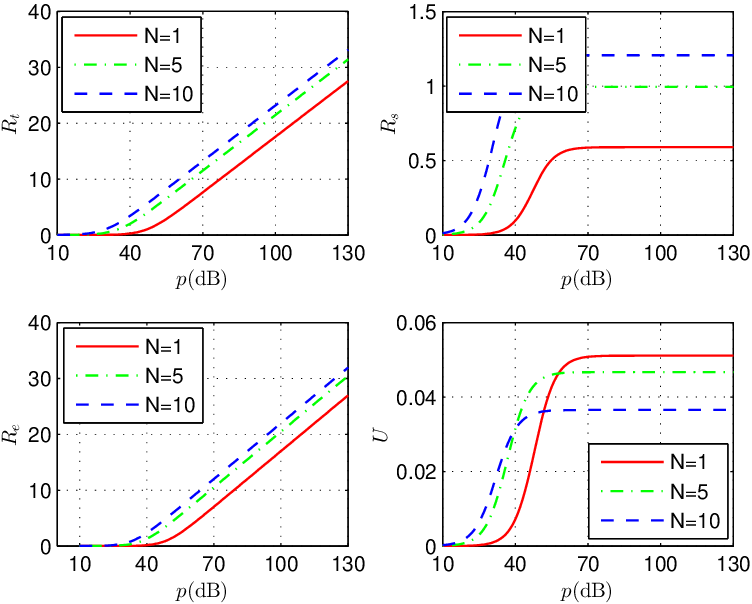}\\
  \caption{Performance of secure transmission under fixed hop-count $N$ versus the transmit power $p$ for the case of NOFT scheme. The system parameters are $L = 50\textrm{m}$, $\epsilon = 0.05$, $\alpha= 3$, $\lambda_e = 10^{-5}$.}
\label{fig:NOFT_P_0_fixN}
\end{figure}
Fig. \ref{fig:NOFT_P_0_fixN} presents the secrecy performance of the system under fixed hop-count $N$ versus the transmit power $p$ for the case of NOFT scheme. Similar trends are observed as in Fig. \ref{fig:OFT_P_0_fixN}, hence the results validate Proposition \ref{coro_p_NOFT}. Looking at the throughput sub-figure, we see that the network with a single hop, i.e., direct transmission, performs very well in this case.

\begin{figure}[!t]
\centering
  \includegraphics[width=9.0cm]{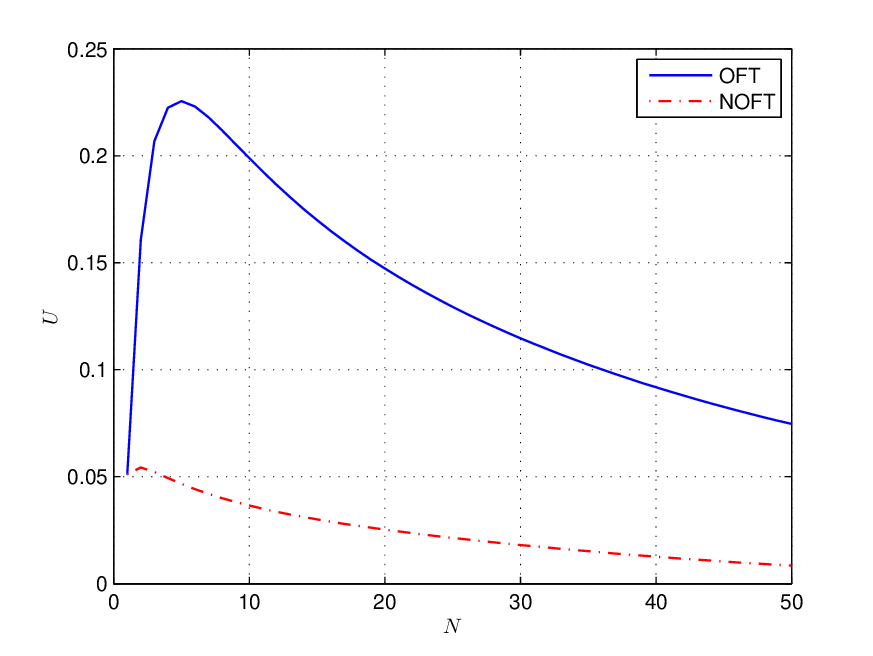}\\
  \caption{The secure transmission throughput of both the OFT and NOFT schemes versus hop-count $N$. The system parameters are $L = 50\textrm{m}$, $\epsilon = 0.05$, $\alpha= 3$, $p =100 \textrm{dB}$, $\lambda_e = 10^{-5}$.}
\label{fig:OFT_NOFT_fixN_compare}
\end{figure}
Fig. \ref{fig:OFT_NOFT_fixN_compare} compares the secure transmission throughput between the OFT and NOFT schemes. As expected, the OFT scheme outperforms the NOFT scheme to a large extent. This highlights the importance of implementing the one-bit feedback for the receiver to inform the transmitter about the current channel condition. By focusing on a single curve in the figure, we can find the optimal hop-count in each scheme: the optimal $N$ for the OFT scheme is 5 while the optimal $N$ for the NOFT scheme is 2.

\subsection{Optimal Number of Hops}

We now explicitly study the optimal hop-count and the resulting secure transmission throughput.

\begin{figure}[!t]
\centering
  \includegraphics[width=7.7cm]{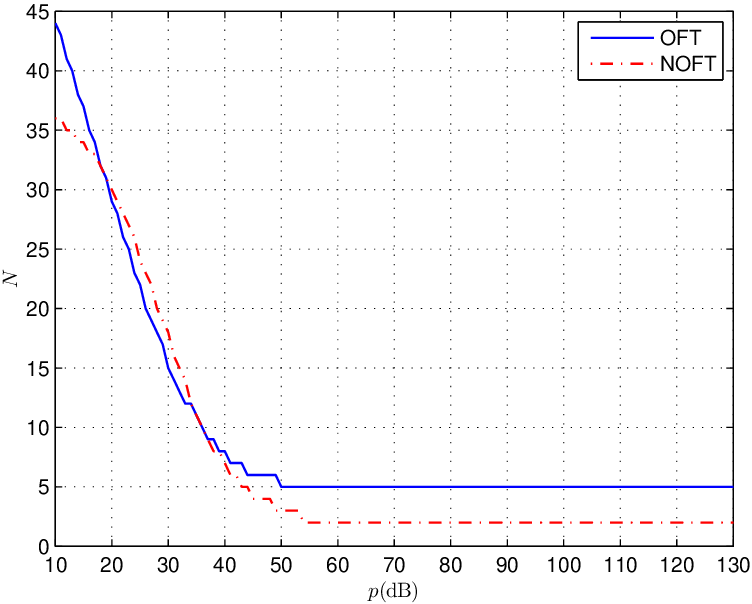}\\
  \caption{The optimal hop-count $N$ versus the transmit power $p$. The system parameters are $L = 50\textrm{m}$, $\epsilon = 0.05$, $\alpha= 3$, $\lambda_e = 10^{-5}$.}
  \label{fig:optimal_N_P_0}
\end{figure}

\begin{figure}[!t]
\centering
  \includegraphics[width=8.0cm]{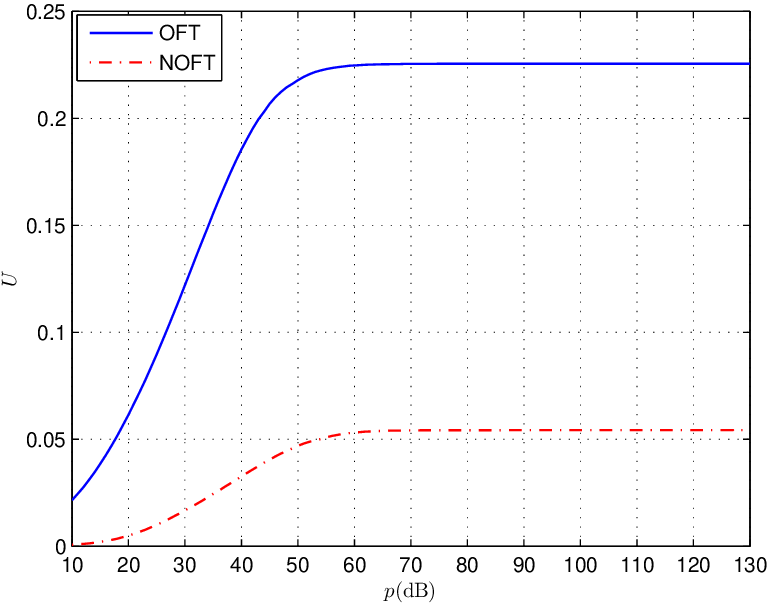}\\
  \caption{The optimal secure transmission throughput $U$ versus the transmit power $p$. The system parameters are $L = 50\textrm{m}$, $\epsilon = 0.05$, $\alpha= 3$, $\lambda_e = 10^{-5}$.}
  \label{fig:optimal_U_P_0}
\end{figure}

Fig. \ref{fig:optimal_N_P_0} presents the optimal hop-count versus the transmit power $p$. As shown in figure, the hop-count of both the OFT and NOFT schemes decrease as the transmission power $p$ increases. This is somewhat intuitive because more transmit power means less hops needed for the end-to-end communication. In addition, more transmit power means better received signal quality at the eavesdroppers for each hop. To avoid the degradation in secrecy, less number of hop (i.e., less number of transmissions) is desired. It is also important to note that the optimal hop-count quickly reaches a constant as $p$ increases. For example, the optimal $N$ reaches and stays at 5 when $p = 50$dB for the OFT scheme and it reaches and stays at 2 when $p = 55$dB for the NOFT scheme. It is worth mentioning that the practical range of $p$ is orders of magnitude higher than 50dB. For example, $p=100$dB is a practical value for sensor-like device with a transmit power of 0dBm and a receiver noise power of -100dBm. Therefore, for practical purposes, we expect that the optimal number of hops are very stable and insensitive to the change in the transmit power. Fig. \ref{fig:optimal_U_P_0} presents the secure transmission throughput achieved using the optimal hop-count. Again, we see that the throughput is constant for practical ranges of $p$. The secrecy performance of OFT scheme is always better than that of the NOFT scheme.

\begin{figure}[!t]
\centering
  \includegraphics[width=8.0cm]{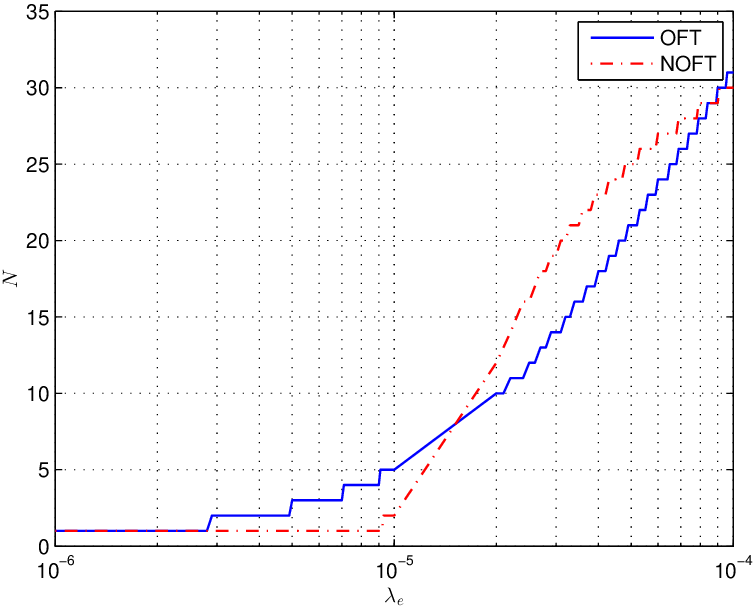}\\
  \caption{The optimal hop-count $N$ versus the density of the eavesdropper $\lambda_e$. The system parameters are $L = 50\textrm{m}$, $\epsilon = 0.05$, $\alpha= 3$, $p = 100 \textrm{dB}$.}
  \label{fig:optimal_N_lambda_e}
\end{figure}

\begin{figure}[!t]
\centering
  \includegraphics[width=8.0cm]{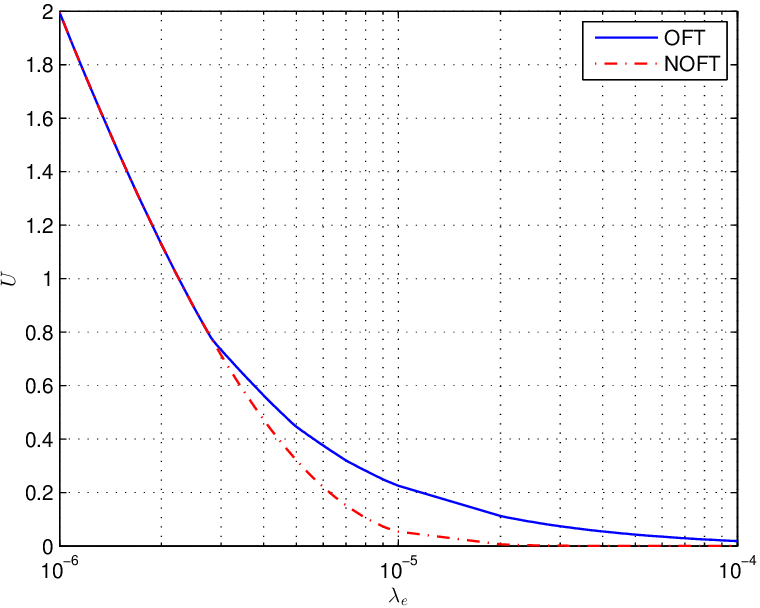}\\
  \caption{The optimal secure transmission throughput $U$ versus the density of the eavesdropper $\lambda_e$. The system parameters are $L = 50\textrm{m}$, $\epsilon = 0.05$, $\alpha= 3$, $p = 100 \textrm{dB}$.}
  \label{fig:optimal_U_lambda_e}
\end{figure}

Fig. \ref{fig:optimal_N_lambda_e} presents the optimal hop-count $N$ versus the density of the eavesdropper $\lambda_e$. As the density of the eavesdropper $\lambda_e$ increases, the optimal number of hops increases. This is not intuitive to understand because more hops means more chances for eavesdropping. The reason why having more hops does not degrade secrecy is due to the fact that more hops reduces the distance of communication among the legitimate nodes which in turn allows a much larger randomness to be added into the wiretap code to fight against eavesdropping while still achieving the same level of communication performance. In short, the benefit of allowing more randomness in the code outweighs the drawback of causing more chances for eavesdropping. Fig. \ref{fig:optimal_U_lambda_e} presents the secure transmission throughput achieved by using the optimal hop-count versus the density of the eavesdropper. When the density of eavesdroppers is very low, the NOFT scheme performs as good as the OFT scheme. When more eavesdroppers are present, the performance of NOFT quickly degrades and becomes much worse than the OFT scheme.

\section{Conclusion}

In this paper, we investigated the secure transmission problem in a linear multihop network with the help of the relays using randomize-and-forward (RaF) relaying strategy in the presence of randomly distributed eavesdroppers. Under an end-to-end secrecy outage probability constraint, we formulated the design problem of maximizing secure transmission throughput by the joint design of the number of hops and the rate parameters of the wiretap code. Both an on-off transmission (OFT) scheme and a non-on-off transmission (NOFT) scheme were studied. Our results give insights into the impact of the system parameters on the secure transmission throughput. For practical ranges of SNR, we observed that the optimal number of hops as well as the secure transmission throughput remains constant and does not change if more transmit power is used. Our results provide design guidelines for determining the best transmission strategy and the best network configuration in a linear multihop network.

\appendices
\section{Proof of Theorem \ref{Theorem_P_so}}\label{appendices_Theorem_P_so}
According to (\ref{snr_e}) and (\ref{P_so_1}), the secrecy outage probability of the $n$th hop for both NOFT and OFT schemes can be computed as
\begin{align}
\nonumber{\mathcal{P}_{\textrm{so}}^{'}} &= {\mathcal{P}}\left ({{\tt SNR}_{{ne}}^\textrm{max}} > \beta_e\right)\\
\nonumber&={\mathcal{P}}\left ({\mathop {\max }\limits_{{e} \in {\Phi _{ne}}} \left\{\frac{{p}{S_{{n}{e}}}}{{|{X_{{n}{e}}}|}^{\alpha}+1} \right\}}> \beta_e \right)\\
&=1-{\mathbb{E}_{\Phi _{ne}}}\left[ {\prod\limits_{{e} \in {\Phi _{ne}} } {\left\{1- \exp\left [-\frac{{\beta_e}{\left({{|{X_{ne}}|}^{\alpha}+1}\right)}}{p} \right]\right\}} } \right].
\label{P_so_3}
\end{align}

According to \cite{Chiu2013}, the probability generating functional (PGFL) for a homogeneous PPP is given as
\begin{align}
{\mathbb{E}_{\Phi _e}}\left[ {\prod\limits_{{e} \in {\Phi _e} } {f\left( {e} \right)} } \right] =
 \exp \left[ { - \lambda_e \int_{{\mathbb{R}^2}} { {1 - f\left( e \right)}\, \text{d}e} } \right].
\label{PGFL}
\end{align}

Using (\ref{PGFL}), (\ref{P_so_3}) can be rewritten as
\begin{align}
{\mathcal{P}_{\textrm{so}}^{'}} = 1-\exp\left[ { - {\lambda_{{e}}}\int_{{\mathbb{R}^2}} {{\exp\left [-\frac{{\beta_e}{\left({{|{X_{ne}}|}^{\alpha}+1}\right)}}{p} \right]} }\, \text{d}e} \right].
\label{P_so_4}
\end{align}

Changing to polar coordinates, (\ref{P_so_4}) can be turned to
\begin{align}
{\mathcal{P}_{\textrm{so}}^{'}} =1-\exp\left[ { - 2\pi\lambda_e \int_{0}^{+\infty} {{\exp\left [-\frac{{\beta_e}{\left({{{r_{e}}}^{\alpha}+1}\right)}}{p} \right]} } {r_e}\, \text{d}r_e} \right].
\label{P_so_5}
\end{align}

Then, (\ref{P_so_5}) can be computed as
\begin{align}
{\mathcal{P}_{\textrm{so}}^{'}} = 1-\exp\left [ -{K_1}\left (\frac{\beta_e}{p}\right)^{-\frac{2}{\alpha}}{}\exp\left [ -\frac{\beta_e}{p} \right ]\right ].
\label{P_so_6}
\end{align}
Replacing ${\mathcal{P}_{\textrm{so}}^{'}}$ with (\ref{P_so_6}) into (\ref{P_so_path_2}), the end-to-end secrecy outage probability of the path can be computed as
\begin{align}
{\mathcal{P}_{\textrm{so}}} = 1-\exp\left [ -{N}{K_1}\left (\frac{\beta_e}{p}\right)^{-\frac{2}{\alpha}}\exp\left [ -\frac{\beta_e}{p} \right ]\right ].
\label{P_so_path_proof_1}
\end{align}

This completes the proof.

\section{Proof of Corollary \ref{coro_fix}}\label{appendices_coro_fix}
We denote the locations of the eavesdroppers in the fixed eavesdroppers case by ${\Phi _{e}}$. Then, the end-to-end secrecy outage probability of the path can be expressed as
\begin{align}
\nonumber{\mathcal{P}_{\textrm{so\_fixed}}} =&\mathcal{P}\left ( \max\limits_{e\in \Phi_{e}}\left \{\max\limits_{n = 1,\ldots,N}\left \{  {\tt SNR}_{{{ne}}}\right\}\right \}> \beta_e \right)\\
\nonumber=&1-\mathcal{P}\left ( \max\limits_{e\in \Phi_{e}}\left \{\max\limits_{n = 1,\ldots,N}\left \{  {\tt SNR}_{{{ne}}}\right\}\right \}< \beta_e \right)\\
=&1-{\mathbb{E}_{\Phi _{e}}}\left[ {\prod\limits_{{e} \in {\Phi _{e}} }\prod\limits_{n = 1}^{N} {\left\{1- \exp\left [-\frac{{{{|{X_{ne}}|}^{\alpha}+1}}}{p/{\beta_e}} \right]\right\}} } \right].
\label{sop_upper_proof}
\end{align}

Using PGFL (\ref{PGFL}), (\ref{sop_upper_proof}) can be rewritten as
\begin{align}
\nonumber{\mathcal{P}_{\textrm{so\_fixed}}} &=1-\exp\Bigg[\\
 -&{ {\lambda_{{e}}}\int_{{\mathbb{R}^2}} {{1-\prod\limits_{n = 1}^{N}\left(1-\exp\left [-\frac{{\beta_e}{\left({{|{X_{ne}}|}^{\alpha}+1}\right)}}{p} \right]\right)} }\, \text{d}e} \Bigg].
\label{sop_upper_proof_1}
\end{align}

\newtheorem{lemma}{Lemma}
\begin{lemma}\label{lemma1}
Let $0<{a_k}\left( {k = 1,2, \ldots,n} \right)<1$ be arbitrary positive constants. For an arbitrary positive integer $n$,
\begin{align}
\prod\limits_{k=1}^n {\left({1-{a_k}}\right)} \ge 1-\sum\limits_{k=1}^n {{a_k}}.
\end{align}
\begin{IEEEproof}
We assume ${f_n} = \prod\limits_{k=1}^n {\left({1-{a_k}}\right)}$ and ${g_n}= 1-\sum\limits_{k=1}^n {{a_k}}$. Then, when $n = 1$, ${f_1} = {g_1} = 1 - {a_1}$. When $n = 2$, ${f_2}-{g_2} = {a_1}{a_2} >0$.

We assume ${f_j}- {g_j} > 0$ when $n = j$. Then, when $n = j+1$, we have
\begin{align}
\nonumber{f_{j+1}}&=\prod\limits_{k=1}^{j+1} {\left({1-{a_k}}\right)}\\
\nonumber&>{g_j}{\left({1-{a_{j+1}}}\right)}\\
\nonumber&={g_{j+1}}+{a_{j+1}}\sum\limits_{k=1}^j {{a_k}}\\
&>{g_{j+1}} .
\end{align}

So we can conclude that $f_n$ is greater than $g_n$ for an arbitrary positive $n \geq 1$.
\end{IEEEproof}
\end{lemma}
\vspace{3ex}

Applying Lemma \ref{lemma1}, we can obtain an upper bound of (\ref{sop_upper_proof_1}) given as
\begin{align}
\nonumber{\mathcal{P}_{\textrm{so\_fixed}}} &\leq 1-\exp\left [
{- {\lambda_{{e}}}\int_{{\mathbb{R}^2}} {{\sum\limits_{n = 1}^{N}\exp\left [-\frac{{{{|{X_{ne}}|}^{\alpha}+1}}}{p/{\beta_e}} \right]} }\, \text{d}e} \right]\\
&=1-\exp\left [ -{N}{K_1}\left (\frac{\beta_e}{p}\right)^{-\frac{2}{\alpha}}\exp\left [ -\frac{\beta_e}{p} \right ]\right ]={\mathcal{P}_{\textrm{so}}}.
\label{sop_upper_proof_2}
\end{align}

This completes the proof.

\section{Proof of Theorem \ref{Theorem_R_t_R_s}}\label{appendices_Theorem_R_t_R_s}
The first derivative of $\mathbb{U}$ w.r.t. $R_t$ is computed as:
\begin{align}
\frac{\mathrm{d} \mathbb{U}}{\mathrm{d} {R_t}} =\frac{\left [ 1-{\ln2}\;{K_2}\;{2^{R_t}}\left ( R_t-R_e \right ) \right ] {\exp\left [ -{K_2}{\left ( 2^{R_t}-1 \right )}\right]}}{N}.
\label{R_t_first_derivative1_1}
\end{align}
Let the first derivative equal to zero, i.e.,
\begin{align}
1-{\ln2}\;{K_2}\;{2^{R_t}}\left ( R_t-R_e \right )  = 0.
\label{R_t_first_derivative1_2}
\end{align}
The value of $R_t$ that satisfies the above equality is given as
\begin{align}
R_t = R_e + \frac{1}{\ln2}\mathbb{W}_0\left (\frac{2^{-R_e}}{K_2} \right ).
\label{R_t_proof}
\end{align}

Also, the second derivative of $\mathbb{U}$ w.r.t. $R_t$ is computed as:
\begin{align}
\nonumber \frac{\mathrm{d}^2 \mathbb{U}}{\mathrm{d} {R_t}^2} =& \frac{{\ln2}\;{K_2}\;2^{R_t}}{N}{\exp\left [ -{K_2}{\left ( 2^{R_t}-1 \right )}\right]}\\
&\times\left [ -2-{\ln2}\left ( R_t-R_e \right )\left ( 2^{R_t}{K_2}-1 \right ) \right ].
\label{R_t_first_derivative2_1}
\end{align}

Replacing $R_t$ with (\ref{R_t_proof}), (\ref{R_t_first_derivative2_1}) can  be rewritten as
\begin{align}
\nonumber \frac{\mathrm{d}^2 \mathbb{U}}{\mathrm{d} {R_t}^2} =& \frac{{\ln2}\;{K_2}\;2^{ R_e + \frac{1}{\ln2}\mathbb{W}_0\left (\frac{2^{-R_e}}{K_2} \right )}}{N}\\
\nonumber&\times{\exp\left [ -{K_2}{\left ( 2^{ R_e + \frac{1}{\ln2}\mathbb{W}_0\left (\frac{2^{-R_e}}{K_2} \right )}-1 \right )}\right]}\\
&\times\left ( -1-\mathbb{W}_0\left (\frac{2^{-R_e}}{K_2} \right ) \right ) < 0.
\label{R_t_first_derivative2_2}
\end{align}

Thus $\mathbb{U}$ is quasi-concave in $R_t$. Since $\frac{1}{\ln2}\mathbb{W}_0\left (\frac{2^{-R_e}}{K_2} \right ) > 0$ is satisfied, we can easily obtain that the constraint (\ref{P2_b}) is also satisfied. Hence, the obtained value of $R_t$ is optimal to maximize $\mathbb{U}$.

Also, combining (\ref{R_e}) and (\ref{R_t_proof}), the optimal value of $R_s$ to maximize $\mathbb{U}$ can be computed as
\begin{align}
R_s = \frac{1}{\ln2}\mathbb{W}_0\left (\frac{2^{-R_e}}{K_2} \right ).
\label{R_s_proof}
\end{align}

This completes the proof.

\section{Proof of Corollary \ref{coro_N}}\label{appendices_coro_N}
From the explicit expression of $R_t^*$, it is not so intuitive to derive the conclusion. In the following, we show the detail procedure of the proof.

Since $\mathbb{W}_0\left (x\right )$ is an increasing function of $x$ when $x>0$, we can easily derive that $R_e$ is also an increasing function of $N$ from the explicit expression of $R_e$ (\ref{R_e}).

\begin{lemma}\label{lemma_N}
Let $0<{C}$ be an arbitrary positive constant. Then $\mathbb{Y} = \mathbb{W}_0\left (\frac{1}{C*z}\right )+\ln\left (z\right )$ is an increasing function of $z$ when $z>0$.

\begin{IEEEproof}
The first derivative of $\mathbb{Y}$ w.r.t. $z$ is computed as:
\begin{align}
\frac{\mathrm{d} \mathbb{Y}}{\mathrm{d} {z}} =\frac{1}{z+z*\mathbb{W}_0\left (\frac{1}{C*z}\right )}>0.
\label{}
\end{align}

So $\mathbb{Y}$ is an increasing function of $z$ when $z>0$.
\end{IEEEproof}
\end{lemma}
\vspace{3ex}

Applying Lemma \ref{lemma_N}, we can obtain that $R_t^*$ is an increasing function of $R_e$ by replacing $z = {2^{R_e}}$ and $C = K_2$ in $\mathbb{Y}$.

Since $K_2$ is an decreasing function of $N$, then we can conclude that $R_t$ is an increasing function of $N$.

When $N\rightarrow +\infty$, $R_e\rightarrow +\infty$ and $K_2 =  \frac{1}{p}$. Then
\begin{align}
R_s^* = {R_t^* - R_e}
= \frac{1}{\ln2}\mathbb{W}_0\left (0 \right ) = 0.
\label{R_s_N_proof}
\end{align}

This completes the proof.

\section{Proof of Corollary \ref{coro_p}}\label{appendices_coro_p}
Replacing $R_e$ and $K_2$, (\ref{R_s}) can be rewritten as
\begin{align}
\nonumber &R_s^* = \\
&\frac{1}{\ln2}\mathbb{W}_0\left (\frac{{p}}{{{\left [ \left ( \frac{L}{N} \right )^\alpha +1 \right ]}\left [{p}\frac{2}{\alpha}\mathbb{W}_0\left (\frac{\alpha}{2} \left [\frac{\ln{\frac{1}{1-\epsilon}}} {{N}{K_1}} \right ]^{-\frac{\alpha}{2}} \right ) +1 \right ]}} \right ).
\label{R_s3}
\end{align}

As ${p}\rightarrow +\infty$, (\ref{R_s3}) can be simplified as
\begin{align}
R_s^* = \frac{1}{\ln2}\mathbb{W}_0\left (\frac{1}{{\frac{2}{\alpha}{\left [ \left ( \frac{L}{N} \right )^\alpha +1 \right ]}\mathbb{W}_0\left (\frac{\alpha}{2} \left [\frac{\ln{\frac{1}{1-\epsilon}}} {{N}{K_1}} \right ]^{-\frac{\alpha}{2}} \right )}} \right ),
\label{R_s4}
\end{align}
which is a constant. With this result, it is easy to see that $R_t^*\rightarrow+\infty$ as ${p}\rightarrow +\infty$ because $R_e\rightarrow+\infty$ as ${p}\rightarrow +\infty$.

Also, the transmission probability ${\mathcal{P}_{t}^{'}}$ can be rewritten as
\begin{align}
{\mathcal{P}_{t}^{'}} = {\exp\left [ -{K_2}{2^{R_s^*+R_e}}-{K_2} \right]}.
\label{P_t_onoff_4}
\end{align}

When ${p}\rightarrow +\infty$,
\begin{align}
\nonumber {K_2}{2^{R_e}} &= \frac{{{\left [ \left ( \frac{L}{N} \right )^\alpha +1 \right ]}\left [{p}\frac{2}{\alpha}\mathbb{W}_0\left (\frac{\alpha}{2} \left [\frac{\ln{\frac{1}{1-\epsilon}}} {{N}{K_1}} \right ]^{-\frac{\alpha}{2}} \right ) +1 \right ]}}{{p}}\\
&= \frac{2}{\alpha}{\left [ \left ( \frac{L}{N} \right )^\alpha +1 \right ]}\mathbb{W}_0\left (\frac{\alpha}{2} \left [\frac{\ln{\frac{1}{1-\epsilon}}} {{N}{K_1}} \right ]^{-\frac{\alpha}{2}} \right ),
\label{throuhput_part2}
\end{align}
\begin{align}
 {K_2} = \frac{\left ( \frac{L}{N} \right )^\alpha +1 }{p} = 0.
\label{throuhput_part3}
\end{align}

Then, (\ref{P_t_onoff_4}) can be simplified as
\begin{align}
\nonumber{\mathcal{P}_{t}^{'}} &= \exp\Bigg [ -\frac{2}{\alpha}{\left [ \left ( \frac{L}{N} \right )^\alpha +1 \right ]}\mathbb{W}_0\left (\frac{\alpha}{2} \left [\frac{\ln{\frac{1}{1-\epsilon}}} {{N}{K_1}} \right ]^{-\frac{\alpha}{2}} \right )\\
&\;\;\times \exp\Bigg [\mathbb{W}_0\Bigg (\frac{1}{{\frac{2}{\alpha}{\left [ \left ( \frac{L}{N} \right )^\alpha +1 \right ]}\mathbb{W}_0\left (\frac{\alpha}{2} \left [\frac{\ln{\frac{1}{1-\epsilon}}} {{N}{K_1}} \right ]^{-\frac{\alpha}{2}} \right )}} \Bigg ) \Bigg] \Bigg].
\label{P_t_onoff_5}
\end{align}

Hence, throughput $\mathbb{U}^*$ can be expressed as
\begin{align}
\nonumber\mathbb{U}^* =&\frac{{1}}{{N}\ln2}\mathbb{W}_0\left (\frac{1}{{\frac{2}{\alpha}{\left [ \left ( \frac{L}{N} \right )^\alpha +1 \right ]}\mathbb{W}_0\left (\frac{\alpha}{2} \left [\frac{\ln{\frac{1}{1-\epsilon}}} {{N}{K_1}} \right ]^{-\frac{\alpha}{2}} \right )}} \right ) \\
\nonumber&\times \exp\Bigg [ -\frac{2}{\alpha}{\left [ \left ( \frac{L}{N} \right )^\alpha +1 \right ]}\mathbb{W}_0\left (\frac{\alpha}{2} \left [\frac{\ln{\frac{1}{1-\epsilon}}} {{N}{K_1}} \right ]^{-\frac{\alpha}{2}} \right )\\
&\times \exp\Bigg [\mathbb{W}_0\Bigg (\frac{1}{{\frac{2}{\alpha}{\left [ \left ( \frac{L}{N} \right )^\alpha +1 \right ]}\mathbb{W}_0\left (\frac{\alpha}{2} \left [\frac{\ln{\frac{1}{1-\epsilon}}} {{N}{K_1}} \right ]^{-\frac{\alpha}{2}} \right )}} \Bigg ) \Bigg] \Bigg].
\label{throuhput_4}
\end{align}

This completes the proof.

\section{Proof of Theorem \ref{Theorem_R_t_R_s_NOFT}} \label{appendices_Theorem_R_t_R_s_NOFT}
According to Theorem \ref{Theorem_R_t_R_s}, we can derive that secure transmission throughput $\mathbb{U}$ is a quasi-concave function of the rate of the transmitted codewords $R_t$ by replacing $K_2$ with $K_4$. Then, the optimal value of $R_t$ and $R_s$ to maximize $\mathbb{U}$ are given as
\begin{align}
R_t^* = R_e + \frac{1}{\ln2}\mathbb{W}_0\left (\frac{2^{-R_e}}{K_4} \right ),
\end{align}
and
\begin{align}
R_s^* = \frac{1}{\ln2}\mathbb{W}_0\left (\frac{2^{-R_e}}{K_4} \right ).
\end{align}

This completes the proof.

\section{Proof of Corollary \ref{coro_p_NOFT}}\label{appendices_coro_p_NOFT}
Replacing $R_e$ and $K_4$, (\ref{R_s_OFT_final}) can be rewritten as
\begin{align}
 \nonumber &R_s^* = \\
 &\frac{1}{\ln2}\mathbb{W}_0\left (\frac{{p}}{{N{\left [ \left ( \frac{L}{N} \right )^\alpha +1 \right ]}\left [{p}\frac{2}{\alpha}\mathbb{W}_0\left (\frac{\alpha}{2} \left [\frac{\ln{\frac{1}{1-\epsilon}}} {{N}{K_1}} \right ]^{-\frac{\alpha}{2}} \right ) +1 \right ]}} \right ).
\label{R_s3_NOFT}
\end{align}

As ${p}\rightarrow +\infty$, (\ref{R_s3_NOFT}) can be simplified as
\begin{align}
R_s^* = \frac{1}{\ln2}\mathbb{W}_0\left (\frac{1}{{\frac{2N}{\alpha}{\left [ \left ( \frac{L}{N} \right )^\alpha +1 \right ]}\mathbb{W}_0\left (\frac{\alpha}{2} \left [\frac{\ln{\frac{1}{1-\epsilon}}} {{N}{K_1}} \right ]^{-\frac{\alpha}{2}} \right )}} \right ),
\label{R_s4_NOFT}
\end{align}
which is a constant. With this result, it is easy to see that $R_t^*\rightarrow+\infty$ as ${p}\rightarrow +\infty$ because $R_e\rightarrow+\infty$ as ${p}\rightarrow +\infty$.

Also, the transmission probability ${\mathcal{P}_{t}^{'}}$ can be rewritten as
\begin{align}
{\mathcal{P}_{t}^{'}} = {\exp\left [ -{K_4}{2^{R_s^*+R_e}}-{K_4} \right]}.
\label{P_t_onoff_4_NOFT}
\end{align}

When ${p}\rightarrow +\infty$,
\begin{align}
\nonumber {K_4}{2^{R_e}} &= \frac{{N{\left [ \left ( \frac{L}{N} \right )^\alpha +1 \right ]}\left [{p}\frac{2}{\alpha}\mathbb{W}_0\left (\frac{\alpha}{2} \left [\frac{\ln{\frac{1}{1-\epsilon}}} {{N}{K_1}} \right ]^{-\frac{\alpha}{2}} \right ) +1 \right ]}}{{p}}\\
&= \frac{2N}{\alpha}{\left [ \left ( \frac{L}{N} \right )^\alpha +1 \right ]}\mathbb{W}_0\left (\frac{\alpha}{2} \left [\frac{\ln{\frac{1}{1-\epsilon}}} {{N}{K_1}} \right ]^{-\frac{\alpha}{2}} \right ),
\label{throuhput_part2_NOFT}
\end{align}
\begin{align}
 {K_4} = \frac{N\left [\left ( \frac{L}{N} \right )^\alpha +1  \right ]}{p} = 0.
\label{throuhput_part3_NOFT}
\end{align}

Then, (\ref{P_t_onoff_4_NOFT}) can be simplified as
\begin{align}
\nonumber{\mathcal{P}_{t}^{'}} &= \exp\Bigg [ -\frac{2N}{\alpha}{\left [ \left ( \frac{L}{N} \right )^\alpha +1 \right ]}\mathbb{W}_0\left (\frac{\alpha}{2} \left [\frac{\ln{\frac{1}{1-\epsilon}}} {{N}{K_1}} \right ]^{-\frac{\alpha}{2}} \right )\\
&\;\;\times \exp\Bigg [\mathbb{W}_0\Bigg (\frac{1}{{\frac{2N}{\alpha}{\left [ \left ( \frac{L}{N} \right )^\alpha +1 \right ]}\mathbb{W}_0\left (\frac{\alpha}{2} \left [\frac{\ln{\frac{1}{1-\epsilon}}} {{N}{K_1}} \right ]^{-\frac{\alpha}{2}} \right )}} \Bigg ) \Bigg] \Bigg].
\label{P_t_onoff_5_NOFT}
\end{align}

Hence, throughput $\mathbb{U}^*$ can be expressed as
\begin{align}
\nonumber\mathbb{U}^* &=\frac{{1}}{{N}\ln2}\mathbb{W}_0\left (\frac{1}{{\frac{2N}{\alpha}{\left [ \left ( \frac{L}{N} \right )^\alpha +1 \right ]}\mathbb{W}_0\left (\frac{\alpha}{2} \left [\frac{\ln{\frac{1}{1-\epsilon}}} {{N}{K_1}} \right ]^{-\frac{\alpha}{2}} \right )}} \right ) \\
\nonumber&\times \exp\Bigg [ -\frac{2N}{\alpha}{\left [ \left ( \frac{L}{N} \right )^\alpha +1 \right ]}\mathbb{W}_0\left (\frac{\alpha}{2} \left [\frac{\ln{\frac{1}{1-\epsilon}}} {{N}{K_1}} \right ]^{-\frac{\alpha}{2}} \right )\\
&\times \exp\Bigg [\mathbb{W}_0\Bigg (\frac{1}{{\frac{2N}{\alpha}{\left [ \left ( \frac{L}{N} \right )^\alpha +1 \right ]}\mathbb{W}_0\left (\frac{\alpha}{2} \left [\frac{\ln{\frac{1}{1-\epsilon}}} {{N}{K_1}} \right ]^{-\frac{\alpha}{2}} \right )}} \Bigg ) \Bigg] \Bigg].
\label{throuhput_4_NOFT}
\end{align}

This completes the proof.

% It is straightforward that $\frac{\mathrm{d} U}{\mathrm{d} {R_t}}>0$ if $R_t < R_e + \frac{1}{\ln2}\mathbb{W}_0\left (\frac{2^{-R_e}}{K_2} \right )$ and $\frac{\mathrm{d} U}{\mathrm{d} {R_t}}\geq 0$ otherwise.
\footnotesize
\bibliographystyle{IEEEtran}
\bibliography{TWC-arXiv}

\end{document}